\documentclass[aps,prl,twocolumn,showpacs]{revtex4}

\usepackage{epsfig}
\begin{document}

\title{Green's function for magnetically incoherent interacting electrons in one dimension}

\author{Gregory A. Fiete$^{1,2,3}$ and Leon Balents$^3$}
\affiliation{$^1$Lyman Laboratory of Physics, Harvard University, Cambridge, MA 02138, USA\\
$^2$Kavli Institute for Theoretical Physics, University of California, Santa Barbara, CA 93106, USA\\
$^3$Physics Department, University of California, Santa Barbara, CA 93106, USA
}

\begin{abstract}
  
  Using a path integral approach and bosonization, we calculate the
  low energy asymptotics of the one particle Green's function for a
  ``magnetically incoherent'' one dimensional strongly interacting
  electron gas at temperatures much greater than the typical exchange
  energy but much lower than the Fermi energy.  The Green's function
  exhibits features reminiscent of spin-charge separation, with
  exponential spatial decay and scaling behavior with interaction
  dependent anomalous exponents inconsistent with any unitary
  conformal field theory.  We compute the tunneling density of states
  at low energies and find that it is a power law in energy with
  exponent $1/(4g)-1$, where $g$ is the Luttinger interaction
  parameter in the charge sector. The underlying physics is made
  transparent by the simplicity of the approach. Our results
  generalize those of Cheianov and Zvonarev [Phys. Rev. Lett. 
{\bf 92}, 176401 (2004)].

\end{abstract}
\date{\today}

\maketitle


Interacting one dimensional electron systems have proven to
be especially rich in their physics, most notably because of the
universal low energy properties that are present when the interactions
are not too strong, the so-called Luttinger Liquid (LL)
state\cite{haldane81} whose existence in nature is now
established\cite{yaro02}.
It is well
known\cite{Voit:rpp94} that the LL state exhibits decoupled spin and
charge degrees of freedom (spin-charge separation) with distinct spin
and charge velocities for the respective collective modes.
Characteristic of the LL state are power-law decays of various
correlation functions, notably the single-particle Green's function,
which is suppressed relative to the na\"ive expectation of Fermi
liquid theory.  As one consequence, the local tunneling density of
states vanishes in a power-law fashion as the chemical potential is
approached.  All low-energy properties of the LL state can be
understood from bosonization, which relates them to correlators in a
simple free boson unitary conformal field theory (CFT).

In the regime of strongly interacting, very low density electrons,
different physics is to be expected when the assumptions of LL theory
break down.  The distinction between low and high densities is often
quantified by the parameter $r_s\equiv (\bar n a_B)^{-1}$ where $\bar
n$ is the average electron density and $a_B$ is the Bohr radius
specific to the material.  When $r_s \gg 1$ the spacing between
electrons is large compared to $a_B$ and the potential energy
dominates the kinetic energy, driving the system towards a Wigner
crystal.  When such strong interactions are present, it becomes
difficult for electrons to exchange their position since they must
effectively tunnel through one another, leading to an exponentially
small\cite{Hausler:zpb96,Matveev:prl04}
(in $r_s$) 
exchange energy, $J$.  It then becomes quite easy to
reach the {\sl magnetically incoherent} regime where the exchange
energy is much less than the temperature: $J \ll T$.  In this Letter
we compute the low energy asymptotics of the 1-d Green's function for
general interactions in the limit $r_s \gg 1$ when the temperature is
still much less than the Fermi energy: $J \ll T \ll E_F$.  We find
that the Green's functions exhibits exponential decay in the spin
sector and power law decay in the charge sector characterized by
interaction dependent {\it anomalous} exponents (which do not
correspond to any unitary conformal field theory).  Our results
generalize the results of Cheianov and
Zvonarev\cite{cheianov03}
(CZ) and the zero field results of 
Berkovich\cite{Berkovich:jpa91} and are obtained in a much
simpler and more physically transparent manner.

Due to breakthroughs in materials technology that allow unprecedented
exploration of clean 1-d quantum
wires\cite{yacoby:sci02}
theoretical interest in such
systems has also been renewed\cite{Nazarov:prl03}
. The
regime $J \ll T \ll E_F$ was recently considered by
Matveev\cite{Matveev:prl04} where he showed there is a drop in the
conductance of a one channel wire from $2e^2/h$ to $e^2/h$ when $J$
drops below $T$.  In the same regime CZ\cite{cheianov03}
have computed the low energy
asymptotics of the one particle Green's function assuming infinitely
strong zero range interactions between electrons. Their tour-de-force
Bethe-ansatz based calculation, however, does not provide clear
insight into the physics.  Moreover, their results are specific to the
special features of the Hamiltonian they considered.  Our calculation
of the 1-d Green's function in the regime $J \ll T \ll E_F$ is
completely general and simple enough to highlight the physical origins
of the non-LL features.

We note that the physics of the $J \ll T \ll E_F$ regime is of broad
and general interest as it may play some role in other systems not
fully understood, such as the two dimensional metal-insulator
transition and the 0.7 anomaly in quantum point contacts. As such, it
is a worthwhile endeavor to elucidate the basic physics of this
regime.

{\it MODEL}: We assume that the electrons are confined to one
dimension and experience predominantly repulsive and {\sl spin
  independent} interactions, i.e.  ones which can be written solely in
terms of the local electron density $n(x)=
\sum_{\sigma=\uparrow,\downarrow} \psi^\dagger_\sigma(x)
\psi^{\vphantom\dagger}_\sigma(x)$.  Here $\psi_\sigma(x)$ is the
field operator for an electron at position $x$ with spin $\sigma$.
In the discussion that follows, we will not require an explicit
Hamiltonian, but rather make use of an effective
low-energy theory that contains renormalized parameters, $v_c$ and
$g$, depending on the microscopic interactions in an unspecified way.
Despite our interest in the magnetically incoherent regime $J\ll T$,
we will see that the essential parameters describing the effect of
interactions may be obtained from the charge sector action of LL
theory which obtains at the lowest temperatures $T\ll J$:
\begin{equation}
S_c=\int dx d\tau \frac{v_c}{2\pi}\left[\frac{1}{2g}(\partial_x \theta)^2+2g(\partial_x \phi)^2 \right]+\frac{i}{\pi}\partial_\tau\phi\partial_x\theta,
\label{eq:action}
\end{equation}
where $v_c$ is the velocity of charge excitations, $g$ is the parameter 
of the low energy theory measuring the strength of
electron interactions 
and the charge fields $\theta$ and $\phi$ are the fields appearing
in the low-temperature bosonized version of the electron operator.
The charge fields are defined via
$\theta_\uparrow+\theta_\downarrow=\theta$ and
$(\phi_\uparrow+\phi_\downarrow)/2=\phi$.  
The $\theta$ field is related to the particle density fluctuations
through the familar relationship
$n(x)=\frac{1}{\pi}\partial_x\theta(x)$ 
and $e^{\pi \phi(x)}$ annihilates/creates a particle at x.
We choose to do all our
calculations in imaginary time, $\tau=it$.  The Green's functions of 
interest (retarded, 
advanced, etc.) can then be computed by the appropriate
analytical continuation to real time.

{\it RESULTS}: We compute the single particle Green's function
\begin{equation}
{\cal G}(x,\tau)=\langle\psi_\uparrow(x,\tau)\psi_\uparrow^\dagger(0,0)\rangle\;,
\label{eq:Green's} 
\end{equation} 
in the limit $J \ll T \ll E_F$ by first averaging over spin
configurations and then computing the low energy charge dynamics using
(\ref{eq:action}).  Due to the spin rotational invariance of
the electron interactions, the Green's function for spin down 
electrons coincides
with (\ref{eq:Green's}).  We find (for $x>0,\tau>0 \rightarrow \infty$),
\begin{equation}
{\cal G}(x,\tau)=\frac{C'e^{-\tilde{k}_F x \frac{\ln 2}{\pi}}}{(x^2+v_c^2\tau^2)^{\Delta_g}}\left(\frac{e^{i(\tilde{k}_F x -\varphi_g^+)}}{v_c \tau-ix}+\frac{e^{-i(\tilde{k}_F x -\varphi_g^-)}}{v_c \tau+ix}\right)
\label{eq:G_asym_x}
\end{equation}
where $C'$ is an undetermined constant (CZ determine it for the
special case of infinite strength zero range interactions in
Ref.~\cite{cheianov03}).  Here we follow CZ and define a {\sl
  spinless} Fermi wavevector
$\tilde{k}_F \equiv \pi \bar n$, where $\bar n$ is the average density
of electrons.
The phases $\varphi_g^\pm$ are given by
\begin{equation}
\varphi^\pm_g(x,\tau)=\frac{\ln 2}{\pi}\left ( g \ln(x^2+v_c^2\tau^2)\pm \frac{1}{2}
\ln \left(\frac{v_c\tau-ix}{v_c\tau+ix}\right) \right)\;.
\label{eq:phases}
\end{equation}
The power law decay of (\ref{eq:G_asym_x})
is characterized by the {\it anomalous} exponent
\begin{equation}
\Delta_g=\frac{1}{8g}+\frac{g}{2}\left(1-\left(\frac{\ln 2}{\pi}\right)^2\right) -\frac{1}{2}\;.
\label{eq:D_g}
\end{equation}

It is clear that the Green's function (\ref{eq:G_asym_x}) does not fit
into the usual LL paradigm of correlation functions with power law
decay because of the exponential decay with distance.  We will show
that this exponential decay is simply a result of spin averaging when
$J \ll T$.  The exponents characterizing the power law decay of the
charge sector are anomalous because for certain values of $g$, $g=1/2$
for example, $\Delta_{g=\frac{1}{2}} = -\frac{1}{4}\left(\frac{\ln
    2}{\pi}\right)^2 <0$, which, if interpreted as arising from CFT,
implies non-unitarity in the charge
sector\cite{cheianov03}.
We will show the anomalous power
law decay of (\ref{eq:G_asym_x}) comes from density fluctuations in
the charge sector after averaging over the spin degrees of freedom.

We also compute the $x=0$ Green's function
\begin{equation}
{\cal G}(0,\tau) \sim \frac{1}{\sqrt{\tau^{\frac{1}{2g}} \ln(\tau)}}\;,
\label{eq:G_asym_us}
\end{equation}
from which the low frequency spectral function can be computed
\begin{equation}
A(\omega) \propto \omega^{\frac{1}{4g} -1}/|\sqrt{\ln(w)}|\;.
\label{eq:dos_us}
\end{equation}

All our results recover the results of CZ in the special case they
considered of infinite repulsive local ($\delta$-function)
interactions, which is just the infinite $U$ limit of the Hubbard
model. The CZ limit corresponds\cite{Schulz:prl90} to $g=1/2$ from
which all their results can be recovered by plugging this value of $g$
into Eqs.~(\ref{eq:phases})-(\ref{eq:dos_us}).(CZ ignore the $\sqrt{\ln(w)}$ factor in (\ref{eq:dos_us}); our results agree exactly at g=1/2.)

{\it DERIVATION OF RESULTS}: We study the single-particle Green's function
for a general (finite range and strength of interactions), strongly
interacting 1-d electron system at finite temperature: 
\begin{equation}
{\cal G}(x,\tau)=\frac{1}{Z}{\rm Tr}\left[e^{-\beta H}
\psi_\uparrow(x,\tau)\psi_\uparrow^\dagger(0,0)\right],\; \tau>0.
\label{eq:G_T} 
\end{equation}
Here $Z\equiv{\rm Tr}\left[e^{-\beta H}\right]$ is the partition
function and $\beta$ is the inverse temperature.  Our results are
based on the {\sl first quantized path integral} representation of
${\cal G}(x,\tau)$ and $Z$ in imaginary time, $\tau =it$.  
As is well-known, $Z$ is
obtained as an integral over world lines (paths) of up and down spin
electrons, with periodic boundary conditions in imaginary time
allowing for permutations of identical particles, i.e. the final
positions of electrons of a given spin polarization must be a
permutation of their initial positions (up and down electrons may not
be permuted as they are not identical).  Each such configuration is
weighted by $(-1)^{P_\uparrow+P_\downarrow} e^{-S_e}$, where $S_e$ is
the Euclidean action describing the ``deformation''
of the world lines and $(-1)^{P_{\sigma}}$ is the sign of the permutation
of spin $\sigma$ electrons.  The numerator of Eq.~(\ref{eq:G_T}) is a
similar integral over paths, with a spin-up electron world line
inserted at $(0,0)$ and another removed at $(x,\tau)$.  To compute the
sign of the permutation, one should treat the path terminating at
$(x,\tau)$ as continuing from $(0,0)$.  

\begin{figure}[h]
\includegraphics[width=1.0\linewidth,clip=]{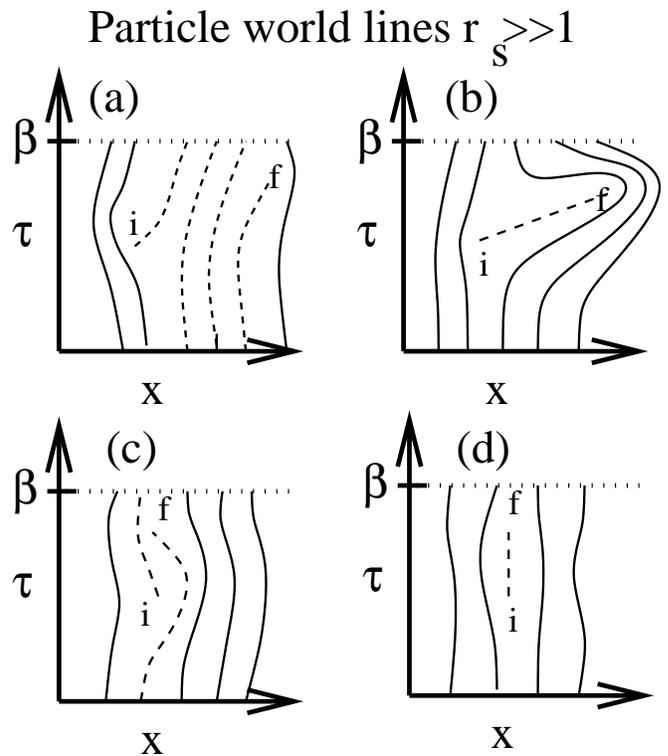}
\caption{\label{fig:worldlines} World lines for a strongly interacting
1-d electron gas at $J<<T$. Particle trajectories
in space and imaginary time are shown as curved lines.  The dashed
lines represent the world line paths for creating a particle and removing it
for large $x=x_f-x_i,\tau=\tau_f-\tau_i$, ((a) and (b)), and for 
$x=0,\tau$ ((c) and (d)). The solid
lines represent trajectories of other particles. Due to the large action
cost associated with the trajectories in (b), at low energies a process like
that shown in (a) where world lines ``wrap around'' from 
$\tau=\beta$ to $\tau=0$  will dominate.  
Such a process, however, requires that all 
dashed world lines have the same spin.  For $J \ll T$ this occurs with
probability $2^{-N}$ as discussed in the text.  Fig.(c) shows a 
contribution to the $k=1$ term of Eq.(\ref{eq:G_0}), however processes
like that shown in (d) dominate and yield the Green's function 
(\ref{eq:G_asym_us}).}
\end{figure}

A crucial element of our analysis is a {\sl non-crossing condition}.
Due to the Pauli principle, world lines of the same spin electrons can
be treated as non-crossing irrespective of interactions.  Moreover,
the condition $J \ll T$ precisely corresponds to the absence of
crossings (exchanges) of opposite spin world lines ($1/J\gg \beta$
gives the typical distance in imaginary time between exchange events).
Physically, this is due to the large Coulomb repulsion of electrons.
Thus the topology of the paths is identical to that of spinless
fermions or infinitely repulsive spinless bosons. Some world line
trajectories are shown in Fig.~\ref{fig:worldlines}.

We choose to compute the trace in Eq.~(\ref{eq:G_T}) by first summing
over all possible spin assignments for each set of world lines.  This
gives an effective weight for the remaining sum over the world line
configurations.  This weight includes two contributions: (1) the
Euclidean action factor $e^{-S_e}$, and (2) a magnetic/statistical
factor obtained from the spin sum and permutation signs, which is
discussed below.  Given the non-crossing topology and spin-independent
interactions, the Euclidean action is expected to be well-approximated
(for $T\ll E_F$) by the low-energy effective form appropriate to
spinless (in general interacting) fermions/bosons, which justifies
Eq.~(\ref{eq:action}).  Moreover, for $J\ll E_F$, we expect this
``charge sector'' action to be identical to the ultimate low-energy
charge action of LL theory valid for $T\ll J$.

Next consider the spin sum.  The limit $J \ll T$ implies that for a
fixed set of world lines all spin
configurations are computed with equal weight.
In order for a particle created at spacetime coordinate $(0,0)$ to be
annihilated at $(x,\tau)$ a large distance away, the world line must
``wrap around'' the imaginary time inverval 0 to $\beta$ a large
number of times. 
From Fig.~\ref{fig:worldlines} it is clear that to stay in the low
energy sector, all intermediate world lines must have the same spin.
Any configuration of world lines without all the intermediate lines
having the same spin would violate the non-crossing condition.  When
the spin average is taken in the trace in (\ref{eq:G_T}), only the
term with all intermediate worldlines having spin up will contribute
to the low energy Green's function.  For $N$ intermediate world lines,
the probability that all $N+1$ spins are the same is $2^{-N}$.
After an electron added at (0,0) reaches $(x,\tau)$, it has been
permuted through $N$ electrons of the same spin and therefore the
Green's function picks up a factor $(-1)^N$.

After spin averaging for $J \ll T$, the Green's function therefore becomes
\begin{equation}
{\cal G}(x,\tau) \sim \langle 2^{-N(x,\tau)} (-1)^{N(x,\tau)} e^{i(\phi(x,\tau)-\phi(0,0))}\rangle\;,
\end{equation}
where now the average is taken at $T \rightarrow 0$ in the charge
sector. The term
$e^{i(\phi(x,\tau)-\phi(0,0))}$ creates a world line
at (0,0) and annihilates it at $(x,\tau)$. (Note that at this stage 
in the calculation the  dynamics after spin averaging become 
effectively spinless as discussed earlier.)  All the effects of
statistics and spin are encapsulated in the first two terms inside the
average.  The number of electrons is related to the $\theta$ field via
\begin{equation}
N(x,\tau)=\bar n x + \frac{1}{\pi}\left(\theta(x,\tau)-\theta(0,0)\right)\;,
\label{eq:N}
\end{equation}
which expressed the number of electrons in a distance $x$ in terms of the
average density and a small fluctuating piece expressed in terms of the 
$\theta$ fields. Using 
Eq.~(\ref{eq:N}),  and writing $(-1)^{N} = {\rm Re}[e^{i\pi
  N}]$, the simplest form correct for integer $N$ (the harmonic approximation violates this) and consistent with the requirement that ${\cal G}(x,\tau)$ be real and even in $x$. The Green's function can now be expressed as ${\cal G}=
{\cal G}_+ + {\cal G}_-$, with ${\cal G}_- = [{\cal G}_+]^*$ and 
\begin{eqnarray}
{\cal G}_+(x,\tau) &\sim&  e^{-\tilde{k}_F x\frac{\ln 2}{\pi}}
e^{i\tilde{k}_Fx}  
\langle e^{-\frac{\ln 2}{\pi}(\theta(x,\tau)-\theta(0,0))}
\nonumber \\
&\times& e^{i(\theta(x,\tau)-\theta(0,0))} e^{i(\phi(x,\tau)-\phi(0,0))}\rangle\;,
\end{eqnarray}
where the first two terms come from the exponentiation of the average
density and we have used $\tilde{k}_F=\pi \bar n$.  This clearly
identifies the exponential decay of the first term as coming from spin
averaging and the oscillatory second term as coming from Fermi
statistics.

We now compute the part of the
Green's function coming from fluctuations in the charge sector 
using the action (\ref{eq:action}).  
Making the definitions, 
$\Phi(x,\tau)=\phi(x,\tau)-\phi(0,0)$ and
$\Theta(x,\tau)=\theta(x,\tau)-\theta(0,0)$, we use
the Gaussian action to move the averages to the exponent,
\begin{eqnarray}
{\cal G}_+(x,\tau) &\sim & e^{-\tilde{k}_Fx \frac{\ln 2}{\pi}}e^{i\tilde{k}_Fx}\langle e^{i\left(1+i\frac{\ln 2}{\pi}\right)\Theta(x,\tau)} e^{i\Phi(x,\tau)}\rangle \nonumber \\
&=&e^{-\tilde{k}_Fx \frac{\ln 2}{\pi}}e^{i\tilde{k}_Fx} e^{-\frac{1}{2}\left( 1+i\frac{\ln 2}{\pi}\right)^2 \langle \Theta^2\rangle}\nonumber \\
& &\times e^{-\frac{1}{2}\langle \Phi^2\rangle } e^{-\left(1+i\frac{\ln 2}{\pi}\right)\langle \Phi \Theta \rangle }\;.
\label{eq:G_bos}
\end{eqnarray}
Standard computations from Eq.~(\ref{eq:action}) give $\langle
\Theta^2\rangle=g \ln(x^2+v_c^2\tau^2)$, $\langle \Phi^2\rangle =
\frac{1}{4g}\ln(x^2+v_c^2\tau^2)$ and $\langle \Phi \Theta \rangle =
\frac{1}{2} \ln\left(\frac{v_c\tau -ix}{v_c\tau+ix}\right)$.
Substituting these values into Eq.~(\ref{eq:G_bos}), the anomalous
exponent given in Eq.(\ref{eq:D_g}) is obtained.  An additional phase
factor comes from the $-\frac{\ln 2}{\pi}(\langle \Theta^2\rangle +
\langle \Phi \Theta \rangle)= -\frac{\ln
  2}{\pi}\left(g\ln
  \left(x^2+v_c^2\tau^2\right) + \frac{1}{2}\ln\left(\frac{v_c\tau-ix}{v_c\tau+ix}\right)\right)$ piece in the exponent.
Combining this with the complex conjugate yields Eq.~(\ref{eq:G_asym_x}).

The preceding calculation brings out the essential physics of the
Green's function (\ref{eq:G_asym_x}): The spin averaging is responsible
for the exponential decay of the Green's function and imposes a
constraint on the world line configurations that contribute to it.
Fermi statistics is responsible for the oscillatory terms.  Treating
Gaussian fluctuations about this constraint in the charge sector
results in the power law decay with generalized anomalous exponents
depending on the interaction parameter $g$.

Having discussed the spatial asymptotics of ${\cal G}(x,\tau)$, for
$x,\tau \rightarrow\infty$, we now turn our attention to ${\cal G}(0,\tau)$ which 
will allow us to compute the low energy tunneling density of states at a 
point. Unlike the situation with $x \rightarrow \infty$,
computing the Green's function at $x=0$ forces us to consider the
discreteness in the number of world lines that may ``bend' in between
(0,0) and $(0,\tau)$:
\begin{eqnarray}
{\cal G}(0,\tau)&\sim& \sum_{k=-\infty}^\infty 2^{-|k|}(-1)^k \langle \delta(N(0,\tau)-k)e^{i(\phi(0,\tau)-\phi(0,0))}\rangle \nonumber \\
\nonumber \\
&\sim& \frac{1}{\sqrt{\tau^{\frac{1}{2g}}{\rm ln}(v_c\tau)}} \sum_{k=-\infty}^\infty 2^{-|k|}(-1)^k e^{-\frac{\pi^2 k^2}{4 g \ln (v_c\tau)}}\;,
\label{eq:G_0}
\end{eqnarray}
where the result (\ref{eq:G_asym_us}) is recovered by noting that the sum 
depends only weakly on $\tau$ and ranges between 2/3 and 1. Fourier 
transforming (\ref{eq:G_0})
into frequency space immediately gives (\ref{eq:dos_us}).

{\it DISCUSSION:} We have computed the low energy asymptotics of the
one particle Green's function ${\cal G}(x,\tau)$ in the limit of 
$r_s \gg 1$ and
$J \ll T \ll E_F$ for arbitrary interactions.  We find the
correlation function does not fit the usual LL form. Instead, the spin
averaging present in the Green's function for $J \ll T$ results in an
exponential decay of the Green's function with distance. The low energy
behavior of the charge sector still shows a power law decay due to
Gaussian fluctuations, but with interaction dependent anomalous
exponents.
The low frequency tunneling density of states (proportional to the
spectral function) also shows interesting behavior depending on $g$:
$A(\omega) \propto \omega^{1/(4g)-1}$, which shows a crossover from a
power law divergence for $g>1/4$, to a pseudo-gap when $g<1/4$. 
The divergence at low
energies can be understood as coming from the infinite spin degeneracy
when $J \ll T$. As the interactions increase ($g$ decreases) the suppression
of the density of states in the charge sector overwhelms the spin degeneracy
to recover the power law suppression familiar in LL theory.  Finally, we note
that Fourier transforming the Green's function (at low frequency) into
momentum space will result in broad peaks of width $\sim \tilde{k}_F$
centered at $k \approx  \pm \tilde{k}_F$.  Note that the
``Fermi momentum'' $\tilde{k}_F$ appearing here differs by a factor of
2 from the usual Fermi momentum: $k_F =\pi \bar{n}/2 = \tilde{k}_F/2$.
Hence, as one moves from $r_s \approx 0$ to $r_s \gg 1$ and the regime
$J \ll T \ll E_F$ is reached, one expects to see delta function-like
peaks present at small $r_s$ to broaden to width $\sim \tilde{k}_F$
and the centers to shift from $\pm \pi \bar n/2$ to $\pm \pi \bar n$
creating a broad double-lobed structure in momentum resolved tunneling
when $r_s \gg 1$ and $J \ll T \ll E_F$.  We hope this work will
inspire new ideas in other systems where the physics discussed here
may play a role.

\acknowledgments

G.A.F. thanks B.I. Halperin for initially bringing Ref
\cite{cheianov03} to his attention and for emphasizing the physical richness of the $J\ll T \ll E_F$ regime. G.A.F. thanks B. I. Halperin,
K . Le Hur, H. Steinberg, Y. Tserkovnyak and A. Yacoby for illuminating
discussions.  This work was supported by DMR-9985255 (L.B.), the
Packard foundation (L.B.),  NSF PHY99-07949 (G.A.F.) and NSF DMR 02-33773 
(G.A.F.).
\vskip -.4 in

\begin{thebibliography}{17}
\expandafter\ifx\csname natexlab\endcsname\relax\def\natexlab#1{#1}\fi
\expandafter\ifx\csname bibnamefont\endcsname\relax
  \def\bibnamefont#1{#1}\fi
\expandafter\ifx\csname bibfnamefont\endcsname\relax
  \def\bibfnamefont#1{#1}\fi
\expandafter\ifx\csname citenamefont\endcsname\relax
  \def\citenamefont#1{#1}\fi
\expandafter\ifx\csname url\endcsname\relax
  \def\url#1{\texttt{#1}}\fi
\expandafter\ifx\csname urlprefix\endcsname\relax\def\urlprefix{URL }\fi
\providecommand{\bibinfo}[2]{#2}
\providecommand{\eprint}[2][]{\url{#2}}

\vskip -.2 in

\bibitem[{\citenamefont{Haldane}(1981)}]{haldane81}
\bibinfo{author}{\bibfnamefont{F.}~\bibnamefont{Haldane}}, \bibinfo{journal}{J.
  Phys. C} \textbf{\bibinfo{volume}{14}}, \bibinfo{pages}{2585}
  (\bibinfo{year}{1981}).

\bibitem[{\citenamefont{Tserkovnyak et~al.}(2002)\citenamefont{Tserkovnyak,
  Halperin, Auslaender, and Yacoby}}]{yaro02}
\bibinfo{author}{\bibfnamefont{Y.}~\bibnamefont{Tserkovnyak}},
\bibnamefont{et~al.}
  \bibinfo{journal}{Phys. Rev. Lett.} \textbf{\bibinfo{volume}{89}},
  \bibinfo{pages}{136805} (\bibinfo{year}{2002}).
%
\bibinfo{author}{\bibfnamefont{H.}~\bibnamefont{Ishii}},
  \bibnamefont{et~al.}, \bibinfo{journal}{Nature}
  \textbf{\bibinfo{volume}{426}}, \bibinfo{pages}{540} (\bibinfo{year}{2003}).
%
\bibinfo{author}{\bibfnamefont{M.}~\bibnamefont{Bockrath}},
\bibnamefont{et~al.},
  \bibinfo{journal}{Nature} \textbf{\bibinfo{volume}{397}},
  \bibinfo{pages}{598} (\bibinfo{year}{1999}).
%
\bibinfo{author}{\bibfnamefont{Z.}~\bibnamefont{Yao}},
\bibnamefont{et~al.},
  \bibinfo{journal}{Nature} \textbf{\bibinfo{volume}{402}},
  \bibinfo{pages}{273} (\bibinfo{year}{1999}).

\bibitem[{\citenamefont{Voit}(1994)}]{Voit:rpp94}
\bibinfo{author}{\bibfnamefont{J.}~\bibnamefont{Voit}}, \bibinfo{journal}{Rep.
  Prog. Phys.} \textbf{\bibinfo{volume}{57}}, \bibinfo{pages}{977}
  (\bibinfo{year}{1994}).

\bibitem[{\citenamefont{H\"ausler}(1996)}]{Hausler:zpb96}
\bibinfo{author}{\bibfnamefont{W.}~\bibnamefont{H\"ausler}},
  \bibinfo{journal}{Z. Phys. B} \textbf{\bibinfo{volume}{99}},
  \bibinfo{pages}{551} (\bibinfo{year}{1996}).

\bibitem[{\citenamefont{Matveev}(2004)}]{Matveev:prl04}
\bibinfo{author}{\bibfnamefont{K.~A.} \bibnamefont{Matveev}},
  \bibinfo{journal}{Phys. Rev. Lett.} \textbf{\bibinfo{volume}{92}},
  \bibinfo{pages}{106801} (\bibinfo{year}{2004}).
%
\bibinfo{note}{K.A. Matveev, cond-mat/0405542.}

\bibitem[{\citenamefont{Cheianov and
  Zvonarev}(2004{\natexlab{a}})}]{cheianov03}
\bibinfo{author}{\bibfnamefont{V.~V.} \bibnamefont{Cheianov}} \bibnamefont{and}
  \bibinfo{author}{\bibfnamefont{M.~B.} \bibnamefont{Zvonarev}},
  \bibinfo{journal}{Phys. Rev. Lett.} \textbf{\bibinfo{volume}{92}},
  \bibinfo{pages}{176401} (\bibinfo{year}{2004}{\natexlab{a}}).
%
\bibinfo{author}{\bibfnamefont{V.~V.} \bibnamefont{Cheianov}} \bibnamefont{and}
  \bibinfo{author}{\bibfnamefont{M.~B.} \bibnamefont{Zvonarev}},
  \bibinfo{journal}{J. Phys. A: Math. Gen.} \textbf{\bibinfo{volume}{37}},
  \bibinfo{pages}{2261} (\bibinfo{year}{2004}{\natexlab{b}}).

\bibitem[{\citenamefont{Berkovich}(1991)}]{Berkovich:jpa91}
\bibinfo{author}{\bibfnamefont{A.}~\bibnamefont{Berkovich}},
  \bibinfo{journal}{J. Phys. A: Math. Gen.} \textbf{\bibinfo{volume}{24}},
  \bibinfo{pages}{1543} (\bibinfo{year}{1991}).

\bibitem[{\citenamefont{Auslaender et~al.}(2002)\citenamefont{Auslaender,
  Yacoby, de~Picciotto, Baldwin, Pfeiffer, and West}}]{yacoby:sci02}
\bibinfo{author}{\bibfnamefont{O.}~\bibnamefont{Auslaender}},
\bibnamefont{et~al.},
  \bibinfo{journal}{Science} \textbf{\bibinfo{volume}{295}},
  \bibinfo{pages}{825} (\bibinfo{year}{2002}).
%
\bibinfo{author}{\bibfnamefont{R.}~\bibnamefont{de~Picciotto}},
\bibnamefont{et~al.},
\bibinfo{journal}{Phys. Rev. Lett.}
  \textbf{\bibinfo{volume}{92}}, \bibinfo{pages}{036805}
  (\bibinfo{year}{2004}).

\bibitem[{\citenamefont{Nazarov and Glazman}(2003)}]{Nazarov:prl03}
\bibinfo{author}{\bibfnamefont{Y.~V.} \bibnamefont{Nazarov}} \bibnamefont{and}
  \bibinfo{author}{\bibfnamefont{L.~I.} \bibnamefont{Glazman}},
  \bibinfo{journal}{Phys. Rev. Lett.} \textbf{\bibinfo{volume}{91}},
  \bibinfo{pages}{126804} (\bibinfo{year}{2003}).
%
\bibinfo{author}{\bibfnamefont{M.}~\bibnamefont{Pustilnik}},
\bibnamefont{et~al.},
\bibinfo{journal}{Phys. Rev. Lett.}
  \textbf{\bibinfo{volume}{91}}, \bibinfo{pages}{126805}
  (\bibinfo{year}{2003}).

\bibitem[{\citenamefont{Schulz}(1990)}]{Schulz:prl90}
\bibinfo{author}{\bibfnamefont{H.~J.} \bibnamefont{Schulz}},
  \bibinfo{journal}{Phys. Rev. Lett.} \textbf{\bibinfo{volume}{64}},
  \bibinfo{pages}{2831} (\bibinfo{year}{1990}).

\end{thebibliography}

 \end{document}